# From the measurement problem to the primitive ontology programme

Michael Esfeld

University of Lausanne, Department of Philosophy

CH-1015 Lausanne, Switzerland

Michael-Andreas.Esfeld@unil.ch, www.michaelesfeld.com

(for Valia Allori, Angelo Bassi, Detlef Dürr and Nino Zanghì (eds.):

*Do wave functions jump?*, book in honour of GianCarlo Ghirardi)

(draft 27 April 2019)

**Abstract**

The paper retraces the development from the measurement problem to the primitive ontology programme. It assesses the contribution of the GRW theory to this programme and discusses the pros and cons of the GRWm matter density ontology and the GRWf flash ontology in comparison to the Bohmian particle ontology. It thereby pursues the evaluation of the proposals for a primitive ontology of quantum physics.

*1.    The measurement problem and the ontology of quantum physics*

This paper retraces the development from the measurement problem to the primitive ontology programme in quantum physics and assesses the contribution of the GRW theory to this programme. This section recalls this development. Section 2 discusses the GRWm matter density field ontology, section 3 the GRWf flash ontology, taking the latter – with Bell (1987, ch. 22) and *pace* Ghirardi et al. (1995) – to be the most important contribution of the GRW theory to the ontology of quantum physics. Section 4 considers the status of the wave function in this context, advocating a wave function realism that does not amount to a dualism of primitive ontology and wave function.

   The measurement problem is the central motivation for collapse theories. As set out by Maudlin (1995) in what has since become the standard formulation, the measurement problem is the fact that the conjunction of the following three propositions is a contradiction:

   1.A The wave-function of a system is *complete*, i.e. the wave-function specifies (directly or indirectly) all of the physical properties of a system.

   1.B The wave-function always evolves in accord with a linear dynamical equation (e.g. the Schrödinger equation).

   1.C Measurements of, e.g., the spin of an electron always (or at least usually) have determinate outcomes, i.e., at the end of the measurement the measuring device is either in a state which indicates spin up (and not down) or spin down (and not up). (Maudlin 1995, p. 7)

Accordingly, there are three possiblities to solve the measurement problem:

1) One can reject (1.A). There is more to the physical systems than what is represented by the wave function. This "more" is traditionally known as "hidden variables" because we do not have experimental access to more than what is extracted from the wave function by means of Born's rule in terms of predictions of measurement outcome statistics. The most prominent theory in this vein is the one going back to de Broglie (1928) and Bohm (1952). It was supported by Bell from the 1960s to the 1980s (see in particular Bell 1987, chs. 4, 7, 17 and 19) and is today known as Bohmian mechanics (see Dürr et al. 1993). According to this theory, quantum systems always have a determinate value of position, which is not tracked by the wave function. It is, however, misleading to call position a "hidden variable", since all that is ever revealed in measurement outcomes are positions and not wave functions (see Bell 1987, p. 166).

2) One can reject (1.B). In this case, one replaces the Schrödinger dynamics with a dynamics that includes the collapse of the wave function. In the textbook presentations of quantum mechanics, going back to von Neumann (1932), this is done in an *ad hoc* manner, with the wave function being supposed to collapse upon measurement. However, neither are measurements a particular type of interaction – over and above gravitation, electromagnetism and the weak and the strong interaction – that requires a specific law, nor are measurement devices natural kinds on a par with electrons, chemical elements, biological species, etc. The theory of Ghirardi, Rimini and Weber (1986) (GRW) improves on this situation by turning the Schrödinger equation into a law for wave function collapse independently of observers, measurements and the like.

3) One can reject (1.C). In this case, one denies that measurements have outcomes. More precisely, all possible outcomes of any measurement are in fact realized, albeit in different branches of the universe, which do not interfere with one another. This solution goes back to Everett (1957). Consequently, every possible future of a person becomes real in the sense that for every possible future of a person, there is a future self that experiences that future. Hence, there is an obvious problem how to account for probabilities in such a theory, and be it subjective probabilities.

However, at the latest since the seminal paper by Allori et al. (2008), it has become clear that describing the situation that we face when it comes to understanding quantum mechanics in terms of these three possibilities to solve the measurement problem is not the whole story. The reason is that these three possible solutions, thus formulated, are concerned only with the dynamics of the wave function – whether the Schrödinger dynamics is complete so that measurements have no outcomes (not 1.C), whether it has to be amended by wave function collapse (not 1.B) or whether there are additional variables that require an additional dynamics (not 1.A). Insofar as they are only concerned with the dynamics, these solutions do not answer the question of ontology, that is, the question of what the wave function refers to – in other words, the question of what the objects in nature are to which the wave function dynamics relates. This is particularly evident in the case of the collapse dynamics: What are the objects that are subject to this dynamics and what does the collapse of the wave function mean for their behaviour?

When considering the possible answers to the question of the ontology of quantum physics, we come to a division into two principled answers. The one possible answer is what is known as wave function monism. In brief, this is the view that the wave function, conceived as physical object, is the physical reality. The most outspoken advocate of this view is Albert (1996 and 2015, chs. 6-7). The wave function is defined on configuration space by contrast to three-dimensional space or four-dimensional space-time. For $N$ particles, configuration space has $3N$ dimensions such that each point of configuration space represents a possible configuration of the *N* particles in three-dimensional space. However, if the wave function on configuration space *is* the physical reality, there is no configuration of anything in another, physical space. Consequently, it is misleading to call this space "configuration space".

Be that as it may, the idea of Albert's wave function monism is, in brief, that the wave function undergoes in the space on which it is defined an evolution such that objects that are functionally equivalent to objects in three-dimensional space or four-dimensional space-time come into existence during this evolution. In order to achieve this aim, Albert is sympathetic to the idea that the wave function undergoes collapse in this space. This shows that the GRW dynamics can go together with an ontology that admits only the wave function of the universe on a very high-dimensional space (known as GRW0). Moreover, in general, the solution that rejects 1.C is associated with wave function monism: the idea then is that the wave function of the universe undergoes an evolution of a division into many branches of the universe through decoherence; objects that realize a functional definition of ordinary physical objects come into existence during this evolution, even if configuration space monism is rejected (see Wallace 2012 for a prominent contemporary defense of Everettian quantum mechanics).

The other possible answer is the primitive ontology programme. According to this answer, the wave function does not provide the ontology of quantum mechanics. It plays only a dynamical role. The ontology consists in a configuration of matter in three-dimensional space or four-dimensional space-time. In order to represent this configuration, one has to add a variable to the wave function, namely a variable for the (primitive) ontology. This stance is associated with the solutions that reject 1.B or 1.A (although the rejection of 1.B can, as mentioned, also go together with a wave function only ontology).

Since Bohm's theory has been cast in its contemporary formulation as Bohmian mechanics (BM) by Dürr et al. (2013, ch. 2, originally published 1992), it has been set out as a primitive ontology theory. Indeed, Dürr et al. introduce this term in their 1992 paper (p. 29 in the reprint Dürr et al. 2013). BM then is based on four postulates: (i) the primitive ontology of point particles in physical space; (ii) a law that describes the evolution of the configuration of point particles of the universe in which the universal wave function figures, known as guiding equation; (iii) the Schrödinger equation as the law that describes the evolution of the wave function; (iv) a typicality or probability measure in terms of the $|\Psi|^2$ density on the level of the universal wave function; from this measure then follows Born's rule for the prediction of measurement outcome statistics for sub-systems described by their own conditional or effective wave-function.

This structure of a physical theory consisting in a primitive ontology, a law for its evolution, a law for the wave function and a procedure to derive probabilities for measurement outcome statistics applies to any quantum theory that admits a configuration of matter in three-dimensional space or four-dimensional space-time and that assigns to the wave function a dynamical role for the evolution of that configuration. Hence, this structure applies independently of what entities the distribution of matter is taken to consist in (particles or something else) and independently of whether the dynamics for the evolution of the wave function is linear or includes collapse. That is why when taking the ontology of quantum physics into account, the available solutions to the measurement problem come down to two ones: either wave function monism or a primitive ontology of a configuration of matter in ordinary space.

## 2. *The ontology of GRW I: matter density field*

Ghirardi et al. (1995) answer the question of ontology by postulating a continuous matter density field that stretches all over space. This theory is known as GRWm, with "m" standing for the matter density variable that is added to the wave function in order to describe the distribution of matter in physical space. Allori et al. characterize the ontology of GRWm in the following manner:

> We have a variable $m(x,t)$ for every point $x \in \mathbb{R}^3$ in space and every time $t$, defined by
>
> $$m(x,t) = \sum_{i=1}^{N} m_i \int_{\mathbf{R}^{3N}} dq_1 \cdots dq_N \delta(q_i - x) |\psi(q_1,\ldots,q_N,t)|^2 . \qquad [(1)]$$
>
> In words, one starts with the $|\psi|^2$-distribution in configuration space $\mathbb{R}^{3N}$, then obtains the marginal distribution of the $i$th degree of freedom $q_i \in \mathbb{R}^3$ by integrating out all other variables $q_j$, $j \neq i$, multiplies by the mass associated with $q_i$, and sums over $i$. ... The field $m(\cdot, t)$ is supposed to be understood as the density of matter in space at time $t$. (Allori et al. 2008, p. 359)

Matter is a continuous, primitive stuff on this view, stretching all over space, in contrast to discrete and thus countable particles. The variable "m" designates matter qua primitive stuff, as again Allori and co-authors make clear:

> Moreover, the matter that we postulate in GRWm and whose density is given by the *m* function does not *ipso facto* have any such properties as mass or charge; it can only assume various levels of density. (Allori et al. 2014, pp. 331-332)

As in BM, mass, charge, etc. are dynamical variables situated on the level of the wave function that come in through their dynamical role for the evolution of matter; they do not designate intrinsic, essential properties of matter (see Brown et al. 1996, Pylkkänen et al. 2015 and Esfeld et al. 2017 for BM). In both BM and GRWm, matter is characterized by position only. In GRWm, matter is primitive stuff that, moreover, admits of different degrees of density at different points or regions of space, with these degrees of density changing in time. Countenancing a variety of degrees of density as a primitive matter of fact is indispensable in GRWm to account for variation: there is matter all over space with its different degrees of density constituting the variation of matter that we perceive. Thus, for instance, a macroscopic object such as rock or a tree is a concentration of the density of matter in a certain region of space at a time.

Hence, the matter density $m(x,t)$ is an additional variable with respect to the wave function. Consequently, an additional law over and above the collapse law for the evolution of the wave function is needed to establish the link between the wave function and its evolution in configuration space on the one hand and the matter density distribution and its evolution in physical space on the other hand. However, in contrast to the particle positions in BM, the matter density variable is fixed by the wave function. Hence, the dynamics of the wave function keeps track of its evolution. Nonetheless, the matter density variable is "hidden" as well in the sense that it is not fully accessible to an observer. Cowan and Tumulka (2016) show that there are facts about the matter density distribution that an observer cannot know. In particular, there are facts about wave function collapse and hence concentration of the matter density in certain points or regions of space that observers cannot measure.

Indeed, Cowan and Tumulka (2016) establish that in any primitive ontology theory, the primitive ontology cannot be fully accessible to an observer, whatever it may be (particles, a matter density field, or something else) and independently of whether or not the primitive ontology and its evolution is specified and kept track of by the wave function and its evolution. The reason is the no-signalling theorem: if the primitive ontology were fully accessible, superluminal signalling would be possible in Bell-type experiments. This fact confirms the conclusion of the previous section, namely that the possible stances in the ontology of quantum physics fall into two camps only, that is, either the primitive ontology camp or the wave function monism camp.

The GRWm ontology of a matter density field in physical space is a primitive ontology that is modelled on the wave function. It goes as far as possible in reading the primitive ontology of matter in physical space off from the wave function in configuration space. Of course, there are no superpositions in the sense of any indeterminacy in physical space. The variable $m(x,t)$ always has one definite value. But it is modelled on the wave function in the sense that it is a continuous field in physical space as the wave function is a field in configuration space. This close connection raises a number of issues that make it doubtful whether the matter density field is the most convincing proposal for an ontology for the GRW dynamics.

On the one hand, the primitive ontology is one of a field in physical space as the wave function is a field in configuration space. On the other hand, the GRW formalism, as any formalism for quantum mechanics, is a formalism in terms of a finite, determinate and thus countable number of discrete particles, providing a dynamics for these particles, in this case in terms of a collapse or spontaneous localization of the wave function of the $i$th particle. This problem can be remediated by switching to a formalism of a continuous spontaneous localization of the wave function as set out in Ghirardi et al. (1990) with the particle labels disappearing (see Egg and Esfeld 2015, section 3). In any case, the fact that a quantum formalism works with a determinate number of particles is no conclusive argument in favour of an ontology of particles.

The more serious problems for the GRWm theory stem from the dynamics. In the first place, the collapse or spontaneous localization of the wave function is achieved mathematically by multiplying the wave function with a Gaussian. Consequently, the collapsed wave function is sharply peaked in a small region of configuration space, but it does not vanish outside that region; it has tails that spread to infinity. This is therefore known as the problem of the tails of the wave function. On its basis, one can object that the GRWm theory does not solve the measurement problem: for instance, in the Schrödinger cat experiment, when the wave function collapses to the outcome dead cat, there then is a high-density dead cat and a low-density live cat. It seems that the low-density cat is just as cat-like (in terms of shape, behaviour, etc.) as the high-density cat, so that there are in fact two cat-shapes in the matter density field, one with a high and another one with a low density. However, one can give arguments against drawing this conclusion so that it is in dispute whether the tails problem implies that the GRWm theory is in trouble solving the measurement problem (Maudlin 2010, pp. 135-138, argues that it fails to do so; see by contrast Wallace 2014, Albert 2015, pp. 150-154, and Egg and Esfeld 2015, section 3).

More importantly, quantum non-locality has unpalatable consequences for the GRWm theory. This is already evident from a simple example that involves only a position superposition, but no entanglement. Consider the thought experiment of one particle in a box that Einstein raised at the Solvay conference in 1927 (see the account of de Broglie 1964, pp. 28-29, and Norsen 2005): the box is split in two halves that are sent in opposite directions, say from Brussels to Paris and Tokyo. When the half-box arriving in Tokyo is opened and found to be empty, the particle is in the half-box in Paris.

The GRWm account of this experiment is this one: the particle is a matter density field that is split in two halves of equal density when the box is divided; these matter densities travel in opposite directions. Upon interaction with a measurement device, one of these matter densities (say the one in Tokyo) vanishes, while the matter density in the other half-box (the one in Paris) increases; the whole matter then is concentrated in one of the half-boxes (in Paris in this case). This is to say that the matter density in one of the half-boxes is delocated instantaneously across an arbitrary distance in physical space upon collapse of the wave function in configuration space. It does not travel with any velocity (see Egg and Esfeld 2014, p. 193). Even if the collapse of the wave function is conceived as a continuous process, the time it takes for the matter density to disappear in one place and to reappear in another place does not depend on the distance between the two places.

This delocation of matter can with good reason be considered as mysterious (as argued in Esfeld and Deckert 2017, pp. 80-81). Such an account is by no means imposed upon us by quantum non-locality (which can in any case be taken to be counter-intuitive). On BM, for instance, the particles always move on continuous trajectories with a finite velocity. The particle trajectories may be correlated with each other independently of their distance in physical space, thus accounting for quantum non-locality; but nothing is ever delocated spontaneously over arbitrary distances in space.

3.   *The ontology of GRW II: flashes*

The problems for the GRWm theory stem from modelling the primitive ontology of matter in physical space on the wave function. However, there is no compelling reason to seek to infer that ontology from the wave function. All that is ever accessible to us in experiments is what is described in terms of the collapse of the wave function in the GRW formalism. This fact suggests the option to consider only the collapses of the wave function as referring to the empirical reality. The resulting primitive ontology then is one of events occurring at space-time points, which are known as flashes, and the theory is known as GRWf.

Such an ontology was proposed by Bell (1987, ch. 22) immediately after the publication of the GRW dynamics (Ghirardi et al. 1986); the term "flashes" was later coined by Tumulka (2006, p. 826). The point-events (flashes) are ephemeral. There are no continuous sequences of them, since they occur only when the wave function collapses. Accordingly, this proposal goes together with the original GRW dynamics of an instantaneous collapse of the wave function. Hence, there is no underdetermination of the primitive ontology of the GRW dynamics: the flash ontology is tied to the formalism of a spontaneous localization of the wave function that occurs instantaneously, whereas the matter density ontology goes with the formalism of a continuous spontaneous localization of the wave function. Again, the distribution of flashes (the collapses) is not entirely accessible to an observer and in that sense "hidden", although the flashes are determined by the dynamics of the wave function (see Cowan and Tumulka 2016). The flash ontology completes the proposals for a primitive ontology of quantum physics. All the central metaphysical conceptions of objects are realized in these proposals: substances as in Bohmian particles, stuff as in the GRWm matter density field and single events has in the GRWf flashes.

The GRWf ontology is not hit by the mentioned objections to the GRWm ontology: there is no problem of particle numbers or discrete objects in the formalism vs. a wave function field, since the particle number indicates the number of flashes that can possibly occur at a time, and nothing more than the events described as wave function collapse is empirically accessible anyway. There is no problem of the tails of the wave function, since only the wave function as sharply peaked around a point in configuration space refers to matter in physical space, namely point-events (flashes). And there is nothing mysterious in the account of quantum non-locality, since there are only flashes whose occurrences are correlated with one another in the case of entanglement; but nothing is ever delocated across space.

The flash ontology certainly is counter-intuitive, because it refuses to recognize any permanent or persisting material objects. There are only ephemeral flashes with a space-time gap between any two of them (cf. the objection of Maudlin 2011, p. 258). However, these gaps may be so tiny that they cannot be perceived. In other words, the flash ontology is a serious candidate for the correct ontology of our empirical world (see also Arntzenius 2012, ch. 3.15). The initial conditions and the subsequent development of the real world can be such that the flashes occur in such a dense manner that our experience of a macroscopic world of permanent objects is accounted for. Bell (1987, p. 205) regards macroscopic objects as galaxies of flashes. By the same token, an ontology of permanent particles as in BM has to make sure that there are enough particles to account for macroscopic objects that appear as continuous, although there is empty space in the sense of a non-vanishing distance between any two particles.

The flash ontology seems to be distinguished as the most parsimonious one. In its light, the Bohmian particle trajectories and the continuous GRWm matter density field look like a surplus structure, since all that to which we have experimental access are the events that are represented by (effective) wave function collapse. However, on closer inspection, it turns out to be questionable how parsimonious the flash ontology really is. The reason is that the flash ontology is committed to a substantival space or space-time in which the flashes occur. Indeed, one can tie the flash ontology to super-substantivalism, that is, the view that an absolute space-time is all there is: space-time flashes occasionally. Again, the flashes are primitive matter, in this case bare particulars: apart from their spatio-temporal location, they do not have any properties. As in the case of the Bohmian particles and the matter density field, mass, charge, etc. are variables that are situated on the level of the wave function and that consist in playing a dynamical role for the evolution of matter instead of being intrinsic, essential properties of material objects.

The commitment to a substantival space-time in the GRWf ontology is a consequence of the fact that there is a gap between any two flashes, while there still is exactly one universe described by exactly one universal wave function and its evolution (instead of each flash being a universe of its own, like a Leibnizean monad). What holds this universe together then is the space-time in which the flashes occur. There may even be times with no flashes at all. Space-time thus takes the position of the substance in which change occurs, in this case change in the number of flashes and the distances between them. This dualism of absolute space-time and matter qua flashes (or qua the flashing vs. the empty space-time) has the consequence that this ontology is not so parsimonious, simple or minimal after all (see Esfeld and Deckert 2017, pp. 83-84).

By contrast, neither GRWm nor BM are committed to a substantival space-time, although both are conveniently formulated in terms of a configuration of matter that is inserted into an absolute space and evolving in an absolute time. The GRWm matter density field can be conceived as a substance that stands on its own, without requiring an underlying space or space-time (cf. the proposal of Rovelli 1997 for an ontology of fields only without a substantival space-time in the context of the general theory of relativity). By the same token, the permanent Bohmian particles can be conceived as being characterized by their relative distances and their change only instead of positions and trajectories in an absolute space. Indeed, a relationalist formulation of BM has been proposed recently (see Dürr et al. 2018).

Moreover, the Bohmian particles can be construed as being individuated by the distance relations in which they stand. This is an important advantage of the Bohmian particle ontology when it comes to the metaphysics of matter: no commitment to a primtive stuff with different degrees of density as a primitive matter of fact or bare particulars (as in the case of the GRWf flashes) is called for, since there are relations available that individuate the particles. Consequently, BM turns out to be in fact close to the stance that is known as ontic structural realism in the metaphysics of science according to which there are no objects with intrinsic essences, a primitive stuff substratum or bare particulars (see Esfeld and Deckert 2017, ch. 2.1).

In any case, we obtain again the result that the dynamical differences – wave function collapse or not – are not central. The central issue is the evaluation of the proposals for a primitive ontology according to criteria such as parsimony or simplicity, coherence and explanatory value (see Esfeld 2014a).

## 4. *The status of the wave function: dynamics, not ontology*

"Primitive ontology" does not signify that there also is a non-primitive or secondary ontology (which would in this case apply to the wave function). "Primitive" signifies that the configuration of matter simply exists. It cannot be derived from anything else, notably not from the wave function. Quite to the contrary, if one admits a primitive ontology of a configuration of matter, the wave function then enters the theory only through its dynamical role, namely its role for the evolution of the configuration of matter. In that sense, it is nomological.

An ontological dualism of a primitive ontology of matter on the one hand and a wave function on the other hand would fall victim to the objection of Brown and Wallace (2005) according to which the primitive ontology would be superfluous in this case, for everything that is empirically accessible is obtained by applying Born's rule to the wave function. Hence, if the wave function exists as an object of its own, it then contains everything to account for the empirical reality – assuming that wave function monism provides indeed a solution to the measurement problem. However, if one denies this, what accounts for measurement outcomes then is the configuration of the primitive ontology. Consequently, the wave function plays only a dynamical role. There is no reason to admit it to the ontology as an object in addition to the configuration of matter.

If the wave function is nomological in the sense that it plays only a dynamical role, one can adopt with respect to it any one of the three main stances that are available in the metaphysics of laws of nature. By the same token as with respect to laws, one can maintain that the wave function is a primitive entity of its own (see Maudlin 2007) or that it is derived from dispositions, powers or structures that are primitive modal entities over and above the primitive ontology (see e.g. Suárez 2015). However, the explanatory value of both these stances is doubtful: they reify the wave function without thereby making progress in the explanation of the evolution of the configuration of matter, because the wave function is defined in terms of its dynamical role for the primitive ontology. Furthermore, since they reify the wave function, these stances are not immune to the objection from Brown and Wallace (2005).

Given that the wave function enters the theory through its dynamical role for the evolution of the configuration of matter, one can adopt the stance that is known as quantum Humeanism: the wave function is reduced to the evolution of the configuration of matter in the sense that the universal wave function is determined by the overall evolution of the configuration of matter of the universe. Given that entire evolution, the universal wave function is fixed. To my knowledge, the first expression of this stance comes from Bell in "The theory of local beables" (1975):

> One of the apparent non-localities of quantum mechanics is the instantaneous, over all space, 'collapse of the wave function' on 'measurement'. But this does not bother us if we do not grant beable status to the wave function. We can regard it simply as a convenient but inessential mathematical device for formulating correlations between experimental procedures and experimental results, i.e., between one set of beables and another. (quoted from the reprint in Bell 1987, p. 53)

"Beable" is Bell's neo-logism for what exists. His words here have an unnecessarily instrumentalistic tone. The decisive point is that the wave function can be seen as being fixed by correlations between sets of beables – that is, the evolution of the configuration of matter (the "local beables" in Bell's sense). Dowker and Herbauts (2005) provide a concrete model of how this can be so in the framework of GRW flashes on a lattice. In recent years, this view has been worked out as a philosophical stance that is inspired by Humean reductionism about laws of nature and therefore known as quantum Humeanism (Miller 2014, Esfeld 2014b, Callender 2015, Bhogal and Perry 2017). On the one hand, this stance avoids at its roots any objection against an ontological dualism of a primitive ontology and a wave function. On the other hand, this stance still is a scientific realism with respect to the wave function, since the wave function is anchored in, more precisely determined by what exists, namely the primitive ontology of a configuration of matter and its evolution (see Esfeld and Deckert 2017, ch. 2.3, for a detailed argument).

In this context, the seminal contribution of the GRW theory is to make clear that the primitive ontology programme is not limited to Bohmian mechanics. In other words, the alternative to wave function monism going back to Everett (1957) is not only Bohm's (1952) theory. There is a spectrum of primitive ontology theories that covers all the traditional metaphysical positions about objects and in which there is a lively debate about the best proposal for a primitive ontology and corresponding dynamics. In particular, the GRW flash theory is a serious contender that enters into competition with the Bohmian particle ontology (although, to my mind, at the end of the day, the arguments for a primitive particle ontology remain compelling – see Esfeld 2014a and Esfeld and Deckert 2017, chs. 2 to 4).

*Acknowledgements*: I'm grateful to Dustin Lazarovici for helpful comments on the draft of this paper.